\documentstyle[prl,aps,multicol]{revtex}
\input{epsf}
\begin{document}
\draft
\title{Schr\"odinger cat animated on a quantum computer}
\author{A. D. Chepelianskii$^{(a,b)}$, and D. L. Shepelyansky$^{(a)}$}    
\address{$^{(a)}$Laboratoire de Physique Quantique, UMR 5626 du CNRS,
Universit\'e Paul Sabatier, 31062 Toulouse Cedex 4, France}
\address{$^{(b)}$Lyc\'ee Pierre de Fermat, Parvis des Jacobins, 31068 
Toulouse Cedex 7, France}
\date{ Received:  2002.2002 }
\maketitle

\begin{abstract} 
We present a quantum algorithm which allows to simulate 
chaos-assisted tunneling in deep 
semiclassical regime on existing quantum computers.
This opens new possibilities for
investigation of  macroscopic quantum tunneling
and realization of  semiclassical Schr\"odinger cat 
oscillations. 
Our numerical studies determine the decoherence rate 
induced by noisy gates for these oscillations and
propose a suitable parameter regime for their
experimental implementation.
\end{abstract} 

\begin{multicols}{2}
\narrowtext

Since 1935  until recently, the metaphor of Schr\"odinger 
cat oscillations between life and death \cite{cat35} 
was considered as a purely theoretical concept.
However, during the last decade such oscillations 
in a quantum limit were observed for two states of a Rydberg atom in 
a quantum cavity \cite{haroche} and 
an experimental evidence was presented 
for a quantum superposition of macroscopically distinct states
in a superconducting quantum interference device (SQUID) \cite{lukens}.
Manifestations of macroscopic quantum tunneling (MQT)
were also observed in magnetization 
experiments with spin-ten molecular magnets $Fe_8$ and $Mn_{12}$
\cite{sessoli,sarachik}. In addition to their fundamental interest
these experiments promise to provide important
applications, e.g.  for solid-state  qubit realization 
\cite{nakamura} and information storage \cite{loss} based on the
Grover algorithm \cite{grover}.

The regime discussed in \cite{cat35} assumes that 
the quantum tunneling takes place for a semiclassical object
with a regular dynamics in two symmetric regions of phase space.
Recently, the investigations of quantum chaos led to an extension
of this concept to the phenomenon of chaos-assisted tunneling
between islands of regular integrable dynamics
separated by a chaotic sea \cite{bohigas}. In this case, due to chaos 
the period $T_u$ of tunneling oscillations becomes very sensitive
to the variation of system parameters and statistical 
description should be used to describe the average distribution
of $T_u$ \cite{bohigas,creagh,kaplan}. This unusual tunneling
is strongly influenced by complex instanton orbits
and scarring effects \cite{creagh,kaplan} and
a chaos enhancement  of tunneling rate by orders of magnitude may be
reached by a small variation of parameters.
The first direct experimental observations of the chaos-assisted tunneling
have been realized recently with cold  \cite{raizen}
and ultracold atoms from a Bose-Einstein condensate \cite{phillips}
thus opening new possibilities for investigation of this
interesting process.
However, quantum tunneling is a very sensitive
phenomenon and experimental studies are complicated by
the decoherence produced by the environment. 
Due to that theoretical investigations of decoherence effects
on MQT were initiated from the very beginning \cite{leggett,leggett1}
and are continued up to now \cite{schon,chudnovsky}.

In this Letter, we show that quantum entanglement and 
quantum computer simulations \cite{steane} can be efficiently used 
to study quantum tunneling in deep semiclassical regime.
We illustrate this on the example of a quantum symplectic
map (double well map) which has a rich phase space 
structure with integrable islands surrounded by 
chaotic sea. Our algorithm has certain similarities 
with the algorithms for the kicked rotator \cite{cat01} and for the 
sawtooth map \cite{benenti01}. It uses the
quantum Fourier transform (QFT) \cite{qft}
and simulates the dynamics of a quantum system 
with $N$ levels in $O((\log_2 N)^4)$ operations per map 
iteration while any known classical algorithm requires
at least $O(N\log_2 N)$ operations. Only one work space 
qubit is required for computations so that $n_q$ qubits
describe a physical system with $N = 2^{n_q-1}$ levels. 
Contrary to \cite{benenti01} the present algorithm 
simulates the quantum dynamics with mixed (chaotic/integrable)
classical  phase space and 
is optimal for the investigation of chaos-assisted 
tunneling in semiclassical regime.
Indeed, while for MQT in molecular magnets \cite{sessoli,sarachik,chudnovsky1}
the effective Planck constant $\hbar$ is inversely proportional 
to the number of spins ($\hbar \propto 1/n_q$) in our algorithm 
$\hbar \propto 2^{-n_q}$. Hence with only 
10 qubits the algorithm allows to study MQT
with the semiclassical parameter being larger by almost 
two orders of magnitude. 
At present the nuclear magnetic resonance (NMR) quantum computer
can perform QFT \cite{qftexp} and operate with up to 
7 qubits \cite{chuang}. This allows to study 
the chaos-assisted quantum tunneling on existing 
NMR based quantum computers and to obtain important 
information about decoherence effects in MQT regime.

In the classical limit the dynamics of our model is described by
the double well map  given by
\begin{equation}
\overline{p}={p} - K d V(x) /dx \;,
\quad
\overline{x}=x+\overline{p} \; \;\; (mod \; 2 \pi) .
\label{clmap}
\end{equation}
Here $(p, x)$ are momentum and coordinate conjugated variables
$(- \pi < x \leq \pi)$, the bars denote the variables
after one map iteration and $V(x)= (x^2-a^2)^2$. In the limit
$K \rightarrow 0$ the map gives the one-dimensional integrable dynamics
in the double well potential $V(x)$ with the frequency of 
small oscillations $\omega_0 = 2 \sqrt{2 K}$. However, for $K > 0$
the higher harmonics of finite step iterations lead to appearance 
of chaotic component surrounding the stability islands
located at $x = \pm a$. A typical example of mixed phase space 
is shown in Fig.1

\begin{figure} 
\centerline{\epsfxsize=7.cm\epsffile{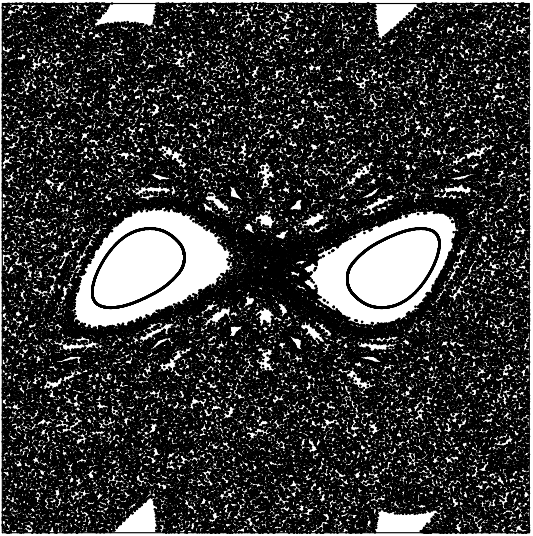}}
\vspace{0.3cm}
\caption{Poincar\'e section for the double well map
(\ref{clmap}) at $K=0.04, a=1.6$, one chaotic and two 
regular orbits are shown in the cell $(-\pi \leq x,p \leq \pi)$.} 
\label{fig1}
\end{figure} 

The quantum evolution on one map iteration is described
by a unitary operator $\hat{U}$ acting on the wave function
$\psi$:
\begin{equation}
\overline{\psi}=\hat{U}\psi =
e^{-i  \hat{p}^2/2\hbar}
e^{-iK V(x)/\hbar}\psi,  
\label{qumap}
\end{equation}
where $\hat{p}=-i \hbar \partial/\partial x$ and $\psi(x+2\pi)=\psi(x)$. 
In the following we take the dimensionless $\hbar = 4 \pi/N$
that corresponds to the case of quantum resonance \cite{izrailev90}
with two classical cells (e.g. as in Fig. 1) on a quantum torus
containing $N$ levels. The semiclassical regime corresponds to 
$\hbar \ll 1$ with discretized momentum $p=\hbar n$
where $n$ is an integer.
The most efficient known classical algorithm simulating the quantum dynamics
(\ref{qumap}) is based on
forward/backward fast Fourier transform (FFT) between $p$ and $x$
representations. For a system with $N$ levels it
requires two FFT and two diagonal multiplications 
in $p$ and $x$ representations and can be performed in
$O(N\log_2(N))$ operations. 

The quantum algorithm simulates 
one map iteration (\ref{qumap}) with $N=2^{(n_q-1)}$ levels
in $O(n_q^4)$ quantum gates operating on $n_q$ qubits with
one qubit used as a work space. The initial wave function 
in $x$ representation is coded in the physical register
with $n_q-1$ qubits in equidistant discrete points $x_m$
$|\psi(x)\rangle = \sum_{m=0}^{N-1} a_m |m\rangle |0\rangle$
with an empty work $n_q-$th qubit. The action of kick 
$U_k=\exp(-iKV(x)/\hbar)$
is diagonal in this representation and the simultaneous multiplication of all
$N$ coefficients can be done in $3 n_q^4$ gate operations.
Indeed, if
$x=\sum_{j=0}^{n_q-2} \alpha_j 2^j$, then
$x^4= \sum_{j_1,j_2,j_3,j_4} \alpha_{j_1} \alpha_{j_2} \alpha_{j_3} \alpha_{j_4} 2^{j_1+j_2+j_3+j_4}$
and  
$e^{-i \beta x^4}=\Pi_{j_1,j_2,j_3,j_4} 
\exp(-i \beta \alpha_{j_1} \alpha_{j_2} \alpha_{j_3} \alpha_{j_4} 
2^{j_1+j_2+j_3+j_4})$ 
with $\alpha_{j_{1,2,3,4}}=0$ or $1$. This step can be performed with
approximately $n_q^4 \;$ 4-qubit gates,
namely control-control-control--phase shift
($C^{(3)}(\beta)$). The gate $C^{(3)}(\beta)$ 
is applied to each group of 4 qubits
and transfers $|1111\rangle$ to 
$\exp(-i \beta 2^{j_1+j_2+j_3+j_4}) |1111\rangle$
keeping other combinations unchanged. Using the work qubit 
the gate $C^{(3)}(\beta)$ can be expressed via
two Toffoli gates $T$ and one control-control--phase shift $C^{(2)}(\beta)$
as $C^{(3)}_{j_1,j_2,j_3,j_4}(\beta) = 
T_{j_1,j_2,w} C^{(2)}_{w,j_3,j_4}(\beta)  T_{j_1,j_2,w}$. 
Here the indices indicate the qubits on which the gates do apply,
$\beta$ notes the rotation angle and $w$ is the work qubit,
which is reset to $0$ after $C^{(2)}(\beta)$.
Thus, the action of kick is expressed via a sequence of standard gates 
used for quantum computations \cite{qft,vedral}. 
Indeed the Toffoli and $C^{(2)}(\beta)$ gates can 
be expressed via one and two qubit gates without 
addition of extra qubits \cite{nielsen}.
The above computation is the most 
\begin{figure} 
\centerline{\epsfxsize=4.2cm\epsffile{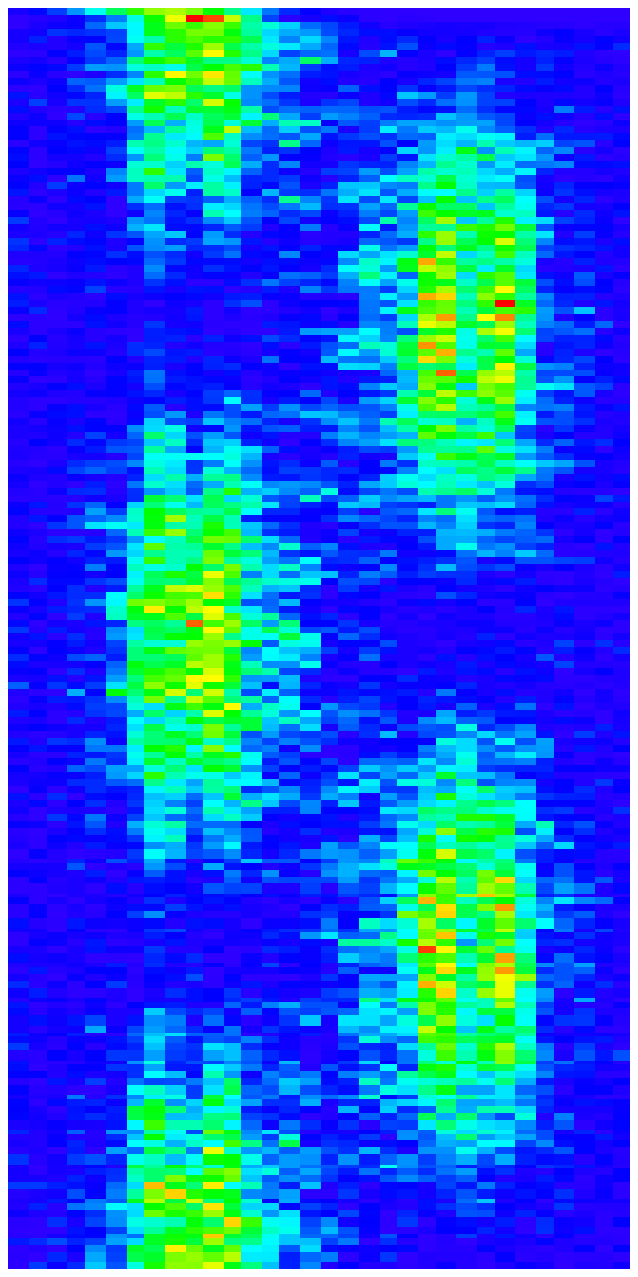}
\hfill\epsfxsize=4.2cm\epsffile{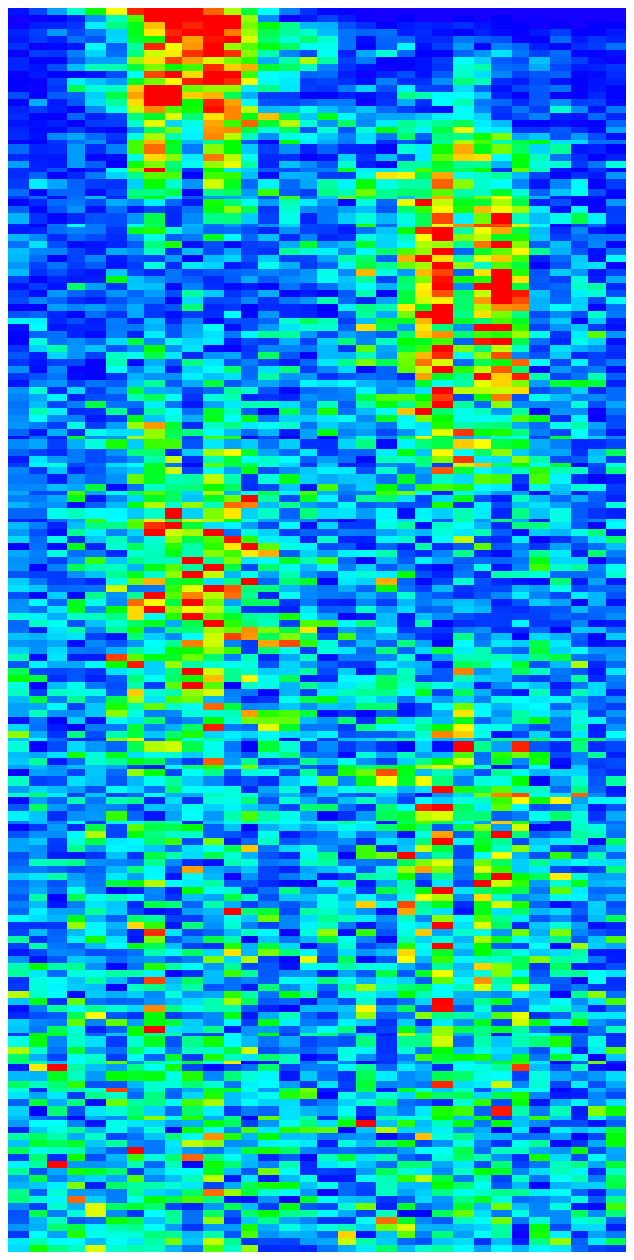}}  
\vspace{.5cm} 
\caption{(color) Time evolution of the Schr\"odinger cat:
probability distribution $W(x)$ at $-\pi \leq x \leq \pi$ is shown
for different number of map iterations $t$, changing along $y$-axis
from $t=0$ (top) to $t=180$ (bottom).
Here as in Fig.1 $K=0.04, a=1.6$ and $\hbar = 4 \pi/N$
with $N=2^{(n_q-1)}$. Quantum computation is done
with $n_q=6$ qubits, ideal perfect gates (left)
and   noisy gates of strength $\epsilon = 0.02$ (right),
and $n_g = 2090$ gates per one map iteration (\ref{qumap}).
At $t=0$ initial coherent packet is
located at $x=-a$. The color is proportional to the density: 
blue for zero and red for maximal density.
}
\label{fig2}       
\end{figure}
\noindent
difficult step in the algorithm and
takes approximately $2 n_q^4$ Toffoli gates
and $n_q^4$ of $C(\beta)$ gates for large $n_q$. The
multiplications by kick phases with lower powers of
$x$ are done in the similar way and require smaller
number of gates. After multiplication by $U_k$ the algorithm 
is similar to the one used in \cite{cat01,benenti01}:
the QFT changes $x$ to $p$ representation in $O(n_q^2)$
operations, the rotation $U_{\hbar} = \exp(-i \hbar n^2/2)$
is realized in $n_q^2$ of control--phase shift gates $C(\beta)$ 
and the backward QFT converts the wave function back to 
initial $x$ representation.

An example of the Schr\"odinger cat oscillations
simulated by this quantum algorithm with ideal gates
in the regime of chaos-assisted tunneling of Fig.1
is shown in Fig.2 (left). The time evolution
of the probability distribution $W(x)$,
integrated over the work qubit, shows clear tunneling
transitions between stability islands of Fig.1.
The same evolution simulated by noisy gates 
is illustrated in Fig.2 (right). 
Noisy gates are modeled by unitary rotations 
by an angle randomly  fluctuating in the interval 
$(-\epsilon/2, \epsilon/2)$ around ideal rotation angle.
This noise introduces an effective decoherence rate 
$\Gamma$ which destroys the tunneling oscillations 
after the time scale $1/\Gamma$.
\begin{figure} 
\centerline{\epsfxsize=8.cm\epsffile{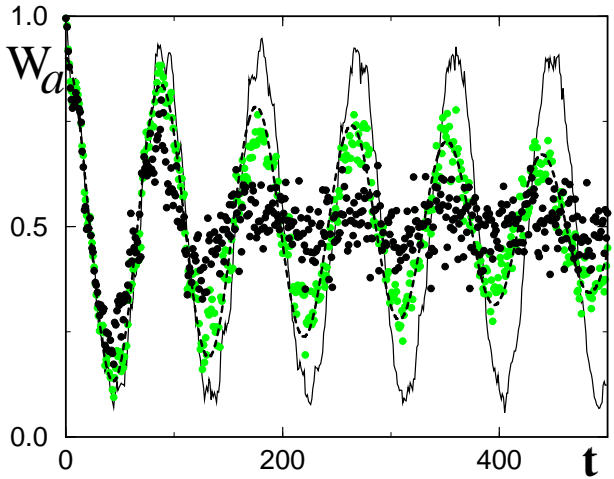}}
\vspace{0.3cm}
\caption{Probability for the Schr\"odinger cat to be alive $W_a$ 
(total probability for $x<0$) as a function of time $t$
for parameters of Fig. 2. 
The time dependence allows to determine
the period of chaos assisted tunneling oscillations 
$T_u = 90$ and their decoherence decay rate  
$\Gamma$. Full curve shows the data without decoherece,
points show the data for noisy gates with $\epsilon =0.01$
(green)
and $\epsilon =0.02$ (black). The fit of 
data gives  $\Gamma = 1.9 \times 10^{-3} $
(dashed curve for $\epsilon  = 0.01$).
}  
\label{fig3}
\end{figure}

To determine the period of tunneling oscillations $T_u$
and their decay rate $\Gamma$ it is convenient
to analyze the time dependence of total probability $W_a(t)$ at $x < 0$
(see Fig.3). The fit 
$W_a(t) - 1/2 \propto e^{-\Gamma t} \cos(2 \pi t/ T_u)$
allows to obtain both $T_u$ and $\Gamma$. We note that
the value of $W_a$ at given $t$ can be obtained 
efficiently from few measurements that enables to determine
$T_u$ and $\Gamma$. Moreover, the values of
$T_u$ and $\Gamma$ are not sensitive to the choice
of initial state at $t=0$. 
As it was discussed in \cite{lloyd} for a similar situation,
this state should only have a sufficiently
large overlap with the coherent state in the center of stability
island. For example, the step distribution, $W(x)=2/N$ for $x<0$
and $W(x)=0$ for $x>0$, which can be prepared efficiently,
gives the same values of $T_u$ and $\Gamma$ as in the case
of the coherent initial state.
\begin{figure} 
\centerline{\epsfxsize=8.cm\epsffile{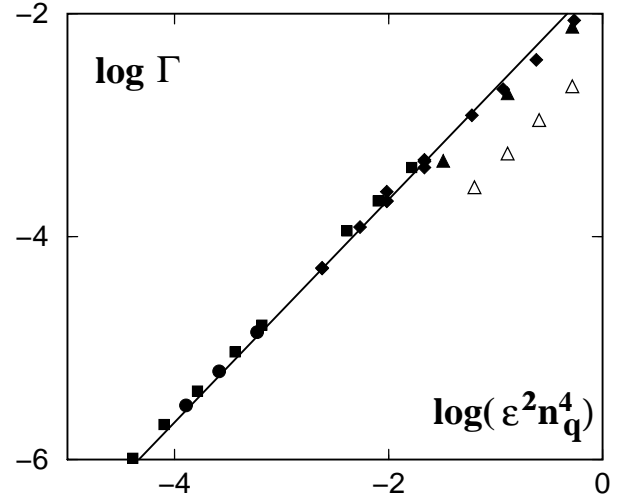}}
\caption{Dependence of 
decoherence rate  of tunneling oscillations $\Gamma$ 
on the strength of gate noise $\epsilon$
for different number of qubits $n_q =$ 6 (full triangles),
7 (diamonds), 8 (squares) and 9 (circles).
The selected map parameters are varied in the range
$1.4 \leq a \leq 1.7$, $0.04 \leq K < 0.06$ at
$\hbar = \pi/2^{(n_q-3)}$ that gave the tunneling period
variation in the range $90 \leq T_u \leq 1.8 \times 10^5$.
The straight line shows the average dependence
(\ref{gamma}). The data with noiseless work qubit
at $n_q=6$ ($K=0.04, a=1.6$) are shown by open triangles. 
Logarithms are decimal. 
}  
\label{fig4}
\end{figure}

The dependence of the decoherence rate on the parameters 
is shown in Fig.4. The variation of $\Gamma$ in a four 
orders of magnitude range is well described by the 
relation:
\begin{equation}
\Gamma = 0.021 \epsilon^2 n_q^4
\label{gamma}
\end{equation}
This relation can be understood using the following 
physical arguments. 
For a given qubit, noise in each unitary gate gives 
a drop of the probability to be directed along the ideal
direction by an amount of $\epsilon^2$. Since at each 
map iteration the number of gates is $n_g \sim n_q^4$ 
the total decay rate is proportional to $ \epsilon^2 n_g$ 
in agreement with Eq.(\ref{gamma}). 
We note that similar estimates for the 
decoherence rate induced by noisy gates 
were also obtained for the Shor algorithm \cite{zurek}.
At the same time the relation (3) is rather simple 
comparing to the decoherence rates discussed for 
MQT in SQUIDs \cite{leggett,leggett1,schon} and molecular magnets
\cite{chudnovsky}. One of the reasons for that 
is that the main step of the algorithm operates 
always with the same work qubit. If the unitary rotations 
of this qubit are noiseless then the decay rate $\Gamma$
is significantly reduced (see Fig.~4).
Hence, the quantum error correcting codes \cite{steane}
applied only to the work qubit can significantly reduce 
the decoherence rate, with a relatively small increase 
of the work space.

Our algorithm allows to obtain interesting results 
about chaos-assisted tunneling even with a small 
number of qubits. For example, the data of Fig.~3 
can be obtained experimentally on the basis of 
techniques applied in \cite{qftexp,chuang}. The main obstacle 
for experimental implementation of this algorithm
is the decoherence. Indeed to observe the Schr\"odinger 
cat oscillations the decoherence time scale $1/\Gamma$ 
should be much larger than the oscillation period $T_u$.
As it is usually the case for semiclassical tunneling, 
the later increases exponentially with the decrease of $\hbar$.
This implies very rapid growth of $T_u$ with $n_q$:
$T_u \propto \exp(S / \hbar) \propto \exp ( 2^{n_q} S / 8\pi)$,
where $S \sim 1$ is a constant related to the classical action.
For example, for $K = 0.04, ~a = 1.6$ the period 
changes from $T_u = 90 ~(n_q = 6)$ to $T_u = 1.68 \times 10^6 ~(n_q = 9).$
Therefore even if the algorithm performs each map
iteration (\ref{qumap}) in polynomial number of gates,
exponential number of map iterations should be done 
to observe tunneling oscillations in deep semiclassical 
regime. Nevertheless, in the regime of chaos-assisted tunneling,
the value of $S$ can be easily varied \cite{bohigas,creagh,kaplan} 
by changing the 
parameters of the map ($K$ and $a$) that allows to obtain
not too large $T_u$ values for $n_q \leq 10$,
e.g. $T_u =305$ for $K=0.3, a=0.5, n_q=10$. Indeed, the value of $S$
can be changed significantly by reducing the size of the stability
islands embedded in the chaotic sea.
In spite of the rapid growth of $T_u$  with the number of qubits $n_q$
the proposed algorithm uses them in an optimal
way in order to reach the minimal effective Plack constant
which scales as $\hbar \propto 2^{-n_q}$.
This situation is qualitatively different comparing to 
SQUIDs \cite{lukens}  where even for a
macroscopic number of particles 
(analogous to $n_q$) the evolution is described by
a Hamiltonian with two levels and effective $\hbar \sim 1$
\cite{schon}.
It also differs from the experiments \cite{raizen,phillips}
where effectively $\hbar \sim 1$  independently of number
of cold atoms.

In summary, our studies show that the Schr\"odinger cat
can be animated in a deep semiclassical regime 
on existing quantum computers \cite{qftexp,chuang} with 
six or more qubits.
Such experiments will give interesting 
information about the nontrivial regime of chaos-assisted 
tunneling in presence of external decoherence. 
They will allow to determine the effective accuracy
of quantum computation,  operability   bounds and
decoherence rates for the first generation of quantum computers.

This work was supported in part by the NSA and ARDA under 
ARO contract No. DAAD19-01-1-0553,
the NSF under Grant No. PHY99-07949 and
the EC RTN contract HPRN-CT-2000-0156.

\end{multicols}

\end{document}